\documentclass[conference, 9pt]{IEEEtran}
\IEEEoverridecommandlockouts

\usepackage{cite}
\usepackage{amsmath,amssymb,amsfonts}
\usepackage{algorithmic}
\usepackage{caption}
\usepackage{subcaption}
\usepackage{graphicx}
\usepackage{textcomp}
\usepackage{xcolor}
\usepackage{tikz,pgf}
\def\BibTeX{{\rm B\kern-.05em{\sc i\kern-.025em b}\kern-.08em
    T\kern-.1667em\lower.7ex\hbox{E}\kern-.125emX}}
\usepackage[per-mode=symbol]{siunitx}
\usepackage[ruled,noline]{algorithm2e}
\usepackage{booktabs}
\usepackage{url}
\renewcommand{\vec}[1]{\mathbf{#1}}

\usepackage{pgfplots}
\usepackage{pgfplotstable}
\usepgfplotslibrary{units}
\usetikzlibrary{pgfplots.statistics, pgfplots.colorbrewer}
\pgfplotsset{compat=1.9}
\makeatletter
\pgfplotsset{
    unit code/.code 2 args=
    \begingroup
    \protected@edef\x{\endgroup\si{#2}}\x
}

\begin{document}

\title{A Zero-Shot Physics-Informed Dictionary Learning Approach for Sound Field Reconstruction
\thanks{This project has received funding from the European Union's Horizon 2020 research and innovation programme under the Marie Sk\text{\l}odowska-Curie grant agreement No. 956962, from KU Leuven internal funds C3/23/056, and from FWO Research Project G0A0424N. This paper reflects only the authors’ views and the Union is not liable for any use that may be made of the contained information.}
\thanks{This work was supported by the Italian Ministry of University and Research (MUR) under the National Recovery and Resilience Plan (NRRP), and by the European Union (EU) under the NextGenerationEU project.}
}
\author{
    \IEEEauthorblockN{Stefano Damiano\IEEEauthorrefmark{1}\IEEEauthorrefmark{2}, Federico Miotello\IEEEauthorrefmark{1}\IEEEauthorrefmark{3}, Mirco Pezzoli\IEEEauthorrefmark{3}, Alberto Bernardini\IEEEauthorrefmark{3},\\Fabio Antonacci\IEEEauthorrefmark{3}, Augusto Sarti\IEEEauthorrefmark{3}, Toon van Waterschoot\IEEEauthorrefmark{2}}

    \IEEEauthorblockA{\IEEEauthorrefmark{1}These authors contributed equally}
    \IEEEauthorblockA{\IEEEauthorrefmark{2}Dept. of Electrical Engineering (ESAT-STADIUS), KU Leuven, Leuven, Belgium
    \\\{stefano.damiano, toon.vanwaterschoot\}@esat.kuleuven.be}
    \IEEEauthorblockA{\IEEEauthorrefmark{3}Dipartimento di Elettronica, Informazione e Bioingegneria, Politecnico di Milano, Milan, Italy
    \\\{federico.miotello, mirco.pezzoli, alberto.bernardini, fabio.antonacci, augusto.sarti\}@polimi.it}

}

\makeatletter
\def\ps@IEEEtitlepagestyle{
  \def\@oddfoot{\mycopyrightnotice}
  \def\@evenfoot{}
}
\def\mycopyrightnotice{
  {\footnotesize
  \begin{minipage}{\textwidth}
  © 20XX IEEE. Personal use of this material is permitted. Permission from IEEE must be obtained for all other uses, in any current or future media, including reprinting/republishing this material for advertising or promotional purposes, creating new collective works, for resale or redistribution to servers or lists, or reuse of any copyrighted component of this work in other works.
  \end{minipage}
  }
}

\maketitle

\begin{abstract}
Sound field reconstruction aims to estimate pressure fields in areas lacking direct measurements.
Existing techniques often rely on strong assumptions or face challenges related to data availability or the explicit modeling of physical properties.
To bridge these gaps, this study introduces a zero-shot, physics-informed dictionary learning approach to perform sound field reconstruction. Our method relies only on a few sparse measurements to learn a dictionary, without the need for additional training data. Moreover, by enforcing the Helmholtz equation during the optimization process, the proposed approach ensures that the reconstructed sound field is represented as a linear combination of a few physically meaningful atoms.
Evaluations on real-world data show that our approach achieves comparable performance to state-of-the-art dictionary learning techniques, with the advantage of requiring only a few observations of the sound field and no training on a dataset.
\end{abstract}


\section{Introduction}\label{sec:introduction}
Sound field reconstruction is a fundamental task in acoustic engineering that focuses on estimating a pressure field in areas where direct measurements are unavailable. Directly measuring the acoustic field across a large volume and the entire audible frequency range requires a significant experimental effort, which is often impractical \cite{hahmann2021a,koyamaMESHRIR2021}. Instead, sound field reconstruction techniques provide a more feasible solution by using a limited set of observations to interpolate and extrapolate the acoustic field~\cite{antonelloRoomImpulseResponse2017,hahmannSpatialReconstructionSound2021,koyamaSparseRepresentationSpatial2019,damianoCompressiveSensingApproach2024,miotelloReconstructionSoundField2024, pezzoliDeepPriorApproach2022,UenoKernelRidgeRegression2018,sundstromImpulseResponseInterpolation2023,pezzoliParametricApproachVirtual2020}. This approach enables a precise acoustic representation of the environment, which is crucial for various applications, such as source localization~\cite{antonelloJointSourceLocalization2018}, sound field control \cite{abe2022amplitude}, auralization \cite{vorlander2020auralization} and active noise cancellation~\cite{brunnstromSpatialActiveNoise2021}.

Various techniques for sound field reconstruction are documented in the literature.
Most approaches are based either on a parametric description of the acoustic scene~\cite{pezzoliParametricApproachVirtual2020,mccormackParametricAmbisonicEncoding2022} or on models derived from the solutions to the wave equation~\cite{uenoSoundFieldRecording2017,UenoKernelRidgeRegression2018}, such as plane waves \cite{jin2015theory}, spherical waves~\cite{borraSoundfieldReconstructionReverberant2019,pezzoliSparsityBased2022}, or equivalent sources~\cite{antonelloRoomImpulseResponse2017,damianoCompressiveSensingApproach2024,koyamaSparseRepresentationSpatial2019}.
While offering robust solutions, these techniques often depend on strong assumptions, such as specific source propagation models, and frequently fall short in accurately modeling real-world scenarios, when the underlying assumptions are no longer valid.

Data-driven approaches try to overcome these limitations by leveraging on training processes carried out on data.
One first class of such methods is based on modeling the acoustic field using neural networks.
Deep learning models, such as convolutional neural networks~\cite{lluis2020sound}, generative adversarial networks~\cite{fernandez-grandeGenerative2023}, and diffusion models~\cite{miotelloReconstructionSoundField2024}, have been successful in learning the acoustic priors from training data and have been exploited to reconstruct pressure fields or address other acoustics-related tasks.
Another class of data-driven methods is constituted by dictionary-learning techniques~\cite{tosicDictionaryLearning2011}. 
These approaches assume that a signal can be represented as a sparse combination of functions, called atoms, learned from observed data, and have demonstrated good performance and generalization capabilities in sound field reconstruction~\cite{hahmannSpatialReconstructionSound2021}. However, the amount of data required to train both deep learning and dictionary learning methods is substantial, and constitutes a bottleneck for their performance.

As a solution, some methods based on neural networks that do not require training on large datasets have also been proposed. For example, the deep prior technique~\cite{ulyanov2018deep,pezzoliDeepPriorApproach2022, miotello2023deep} has shown effectiveness in modeling the acoustic prior of sound fields without needing training data. However, these methods are often unable to outperform data-driven techniques and may struggle to explicitly model the physical properties of sound fields.
Similarly, physics-informed neural networks have been recently introduced to combine the explicit modeling capabilities of traditional model-based techniques with the flexibility of neural networks, while minimizing the amount of data required. These models have been effective in various acoustic problems by incorporating the physical equations governing sound propagation, such as the wave equation, to improve the accuracy~\cite{koyama2024physics, sundstrom2024sound,ribeiro2024sound, miotello2024physics, olivieri2024physics,morena2024reconstruction}.
While they produce physically meaningful solutions, they often lack interpretability.

To improve interpretability and to further integrate model-based techniques with data-driven methods, we propose a zero-shot physics-informed dictionary learning approach for sound field reconstruction.
As a zero-shot technique, our solution only relies on few sparse measurements of a pressure field to learn a dictionary of atoms, with no additional data used for model training. Similarly to \cite{lu2021physics, tetaliWavePhysicsInformed2019}, the physics-informed nature of our approach guides the dictionary learning process by encouraging the atoms to satisfy the underlying governing partial differential equation, i.e., the Helmholtz equation. By enforcing this acoustic prior information, the atoms do not need to be learned from additional training data other than the few observed measurements.
This way, we are able to represent the sound field as a linear combination of a few physically meaningful atoms.
Extensive evaluation on real-world data demonstrates the effectiveness of our method compared to a baseline based on a dictionary of Bessel functions and the state-of-the-art dictionary learning technique~\cite{hahmannSpatialReconstructionSound2021}, that leverages on locally learned and data-driven functions.

\section{Problem Formulation}
\label{sec:problem}
\subsection{Problem statement}
Let us consider a source-free, two-dimensional region of space $\Omega$ lying on a horizontal plane in a reverberant room. The complex-valued
frequency-domain acoustic pressure at position $\vec{r} \in \Omega$ and angular frequency $\omega=2\pi f$, where $f$ denotes the temporal frequency, is expressed by $p(\vec{r},\omega)$. We sample $\Omega$ at $N^2$ points forming a square $N\times N$ grid at positions $\vec{r}_n$, with equidistant points along both directions and inter-point distance $h$. The pressure values $\vec{p} \in \mathbb{C}^{N^2}$ at the $N^2$ grid points can then be expressed as $\vec{p} = [p(\vec{r}_1, \omega), p(\vec{r}_2, \omega), \ldots, p(\vec{r}_{N^2}, \omega)]^\top$. We assume that only a limited number $M < N^2$ of measurements $\tilde{\vec{p}} \in \mathbb{C}^M$ is available, with $\tilde{\vec{p}}$ being a subset of $\vec{p}$. Our goal is to recover the complete set of pressure values $\vec{p}$ from the observations $\tilde{\vec{p}}$, using a zero-shot physics-informed dictionary learning algorithm.
\subsection{Background}
In dictionary-based methods for sound field reconstruction, the acoustic pressure $\tilde{\vec{p}}$ at a sparse set of measurement positions can be modelled as~\cite{hahmannSpatialReconstructionSound2021}
\begin{equation}\label{eq:signal_model}
    \tilde{\vec{p}} = \tilde{\vec{D}} \vec{x} + \tilde{\vec{e}}\,,
\end{equation}
where $\tilde{\vec{D}} \in \mathbb{C}^{M\times L}$ is a dictionary of $L$ position-dependent basis functions evaluated at $M$ arbitrary positions on the grid $\tilde{\vec{r}}_m, 1\leq m \leq M$, $\vec{x} \in \mathbb{C}^L$ is the coefficients vector containing the unknown weights of the atoms, and $\tilde{\vec{e}}$ denotes the measurement noise. The atoms in the dictionary $\tilde{\vec{D}}$ typically contain functions that represent solutions to the Helmholtz equation, e.g. plane waves~\cite{antonelloRoomImpulseResponse2017}, Green's functions~\cite{damianoCompressiveSensingApproach2024}, or, alternatively, can be learned from data~\cite{hahmannSpatialReconstructionSound2021}. In a first analysis stage, and given a known $\tilde{\vec{D}}$ dictionary, the weights $\vec{x}$ can be retrieved by solving the regularized inverse problem~\cite{hahmannSpatialReconstructionSound2021}
\begin{equation}\label{eq:regularized_inverse_problem}
    \hat{\vec{x}} = \arg\min_\vec{x} \lVert \tilde{\vec{p}} - \tilde{\vec{D}}\vec{x} \rVert_2^2 + \alpha \lVert \vec{x} \rVert_p^p\,,
\end{equation}
where the $\ell_p$ norm is defined as
\begin{equation}
    \lVert \vec{x} \rVert_p^p = \sum_{l=1}^L \lvert x_l \rvert^p \,.
\end{equation}
Well-known approaches use the $\ell_0$ or $\ell_1$ norms to promote sparsity in the solution~\cite{verburgReconstructionSoundField2018,antonelloRoomImpulseResponse2017,koyamaSparseRepresentationSpatial2019,damianoCompressiveSensingApproach2024}, assuming $\tilde{\vec{D}}$ to be overcomplete. In a second synthesis stage, $\vec{p}$ can then be reconstructed at all $N^2$ positions through
\begin{equation}\label{eq:reconstruction_equation}
    \vec{p} = \vec{D}\hat{\vec{x}}\,
\end{equation}
where $\vec{D} \in \mathbb{C}^{N^2\times L}$ contains the position-dependent basis functions evaluated at $\vec{r}_n, 1\leq n \leq N^2$.

Dictionary learning relies on the assumption that a signal can be expressed as a sparse combination of functions, called atoms, learned from observed data~\cite{tosicDictionaryLearning2011}. As opposed to defining a closed-form dictionary $\vec{D}$ of basis functions based on some model of the sound field, the dictionary learning framework has been adopted in~\cite{hahmannSpatialReconstructionSound2021} to learn the dictionary atoms from an extensive set of training data.
In the training phase, the overcomplete dictionary $\vec{D}$ and the corresponding training weights $\vec{x}_\text{tr}$ are jointly obtained from a set of training signals $\vec{p}_\text{tr}$ via the optimization problem
\begin{equation}\label{eq:traditional_dictionary_learning}
   \hat{\vec{D}},\hat{\vec{x}}_\text{tr} = \arg\min_{\vec{D},\vec{x}_\text{tr}} \lVert \vec{p}_\text{tr} - \vec{D}\vec{x}_\text{tr} \rVert_2^2 + \alpha \lVert \vec{x}_\text{tr} \rVert_1\,,
\end{equation}
solved by means of alternate optimization algorithms~\cite{aharonK-SVD2006, marialOnlineLearningMatrix2010}. The learned dictionary is then used to reconstruct the sound field in the evaluation setting using the available observations $\tilde{\vec{p}}$, by solving the inverse problem~\eqref{eq:regularized_inverse_problem} with a sparsity-promoting norm to retrieve the weights $\hat{\vec{x}}$, finally plugged in~\eqref{eq:reconstruction_equation} to obtain $\vec{p}$.

Inspired by image-processing techniques~\cite{brucksteinFromSparseSolutions2009}, in~\cite{hahmannSpatialReconstructionSound2021} patch-based processing is adopted to analyze the pressure field. The global domain $\Omega$ is divided in overlapping subdomains, where the pressure field is analyzed locally; the global sound field is then reassembled from the local reconstructions. 
Dictionary learning is adopted to jointly estimate a dictionary and the weights for each analyzed local patch. The model is pre-trained using real-world data recorded in a training room, and the dictionary is then used to effectively reconstruct the sound field in an unseen test room, where only a sparse set of measurements is available, by solving the $\ell_1$-regularized inverse problem~\eqref{eq:regularized_inverse_problem}. Note that the pre-training requires data recorded at all $N^2$ grid points in the training room in order to obtain the dictionary that will be used to reconstruct the sound field in the test room. Despite its effectiveness, this approach requires a pre-training stage and, thus, training data. Moreover, since the dictionary atoms are entirely learned from data, they might lack a physical meaning as they do not necessarily follow the governing Helmholtz equation.
\section{Proposed Method}\label{sec:method}
To promote the retrieval of a physically meaningful dictionary, we introduce a zero-shot physics-informed dictionary learning algorithm that, differently from~\cite{hahmannSpatialReconstructionSound2021}, operates at a global level (i.e., without considering local patches) and does not require pre-training. 
Inspired by~\cite{tetaliWavePhysicsInformed2019}, we encourage structure in the  dictionary by enforcing each atom to satisfy the Helmholtz equation for a certain angular frequency $\omega$. 
The Helmholtz equation can be expressed for each dictionary atom (i.e., column) $\vec{d}_l = \vec{D}_{(:,l)} \in \mathbb{C}^{N^2}$, with $ l=1,\ldots,L$, as~\cite{williamsFourierAcousticsSound1999}
\begin{equation}\label{eq:helmholtz_eq}
    (\nabla^2 + k^2) \vec{d}_l = \vec{0}\,,
\end{equation}
where $k=\omega/c$ denotes the wave number and $c=\SI{343}{\meter\per\second}$ the speed of sound in air. If the distance $h$ between adjacent grid points is sufficiently small, the Helmholtz equation can be discretized using the finite difference method (FDM)~\cite{tetaliWavePhysicsInformed2019}. To this end, we first define the second-order difference matrix $\vec{H}(k) \in \mathbb{R}^{N^2 \times N^2}
$, a symmetric, 5-diagonal Toeplitz matrix whose upper row $\vec{H}_{(1,:)}(k)$ is defined as
\begin{equation}
\vec{H}_{(1,:)}(k) = [\overbrace{-4+k^2h^2 \quad 1 \quad 0 \quad \ldots \quad 1}^{N} \quad 0 \quad \ldots \quad 0]\,.
\end{equation}
Note that the main diagonal elements of $\vec{H}$ depend on frequency through the wave number $k$. Then,~\eqref{eq:helmholtz_eq} can be expressed as 
\begin{equation}\label{eq:discrete_helmholtz}
    \vec{H}(k)\vec{d}_l = \vec{0}\,.
\end{equation}
With this definition, and considering a set of $L$  angular frequencies $\omega_l$, we modify~\eqref{eq:traditional_dictionary_learning} to introduce a regularization term that enforces the atoms of $\vec{D}$ to be solutions of the Helmholtz equation as
\begin{equation}\label{eq:dl_problem}
\begin{aligned}
    \hat{\vec{D}},\hat{\vec{x}} &= \arg\min_{\vec{D,x}} \lVert \tilde{\vec{p}} - \tilde{\vec{D}}\vec{x} \rVert_2^2 + \\&+ \alpha \lVert \vec{x} \rVert_1 
    + \beta \sum_{l=1}^L \lVert \vec{H}(\frac{\omega_l}{c})\vec{d}_l \rVert_2^2\,.
\end{aligned}
\end{equation}
Note that $\tilde{\vec{D}}$ is a subset of $\vec{D}$, containing the $M$ rows corresponding to the positions where measurements are available. Without the Helmholtz regularization term, only these rows would be updated and, thus, the learned dictionary would be meaningless and lead to a poor reconstruction. The Helmholtz regularization term encourages each atom of the \textit{full} dictionary $\vec{D}$ to represent a solution of the Helmholtz equation at a specific frequency $\omega_l$: this way, the entire dictionary gets updated, capturing a wideband representation of the sound field. If all frequencies $\omega_l$ fall within a given range, the learned dictionary can be effectively used to reconstruct the sound field at unseen frequencies within the same range,
as will be discussed in Sec.~\ref{sec:evaluation}. Minimizing \eqref{eq:dl_problem} ensures that the solution remains sparse in terms of the coefficients $\vec{x}$.
Since both $\vec{D}$ and $\vec{x}$ are unknown, the solution to the nonconvex problem in \eqref{eq:dl_problem} is found using an alternating optimization approach \cite{hahmannSpatialReconstructionSound2021}. This process consists of two separate update steps.
First, in the sparse coding step, the optimal coefficients $\vec{x}$ are determined for a fixed dictionary $\vec{D}$ according to \eqref{eq:regularized_inverse_problem}. Then, in the dictionary update step, each atom of the dictionary $\vec{D}$ is optimized by minimizing the discrepancy between the measured pressure field and the reconstructed field at positions $\tilde{\vec{r}}_m$. The two steps are repeated until a maximum number of iterations $I$ is reached. The algorithm is summarized in Algorithm~\ref{algo:proposed_dictionary_learning}.
\begin{algorithm}[t]
    \caption{The Helmholtz-regularized dictionary learning algorithm}
    \label{algo:proposed_dictionary_learning}
        Initialize $\vec{D}$, $\vec{x}$\\
        Set $\alpha, \beta, L, I$\\
        \For{$i=0$ \textup{to} $I$}
        {
        \textit{Sparse coding:} \\
        $\hat{\vec{x}} = \arg\min_\vec{x} \lVert \tilde{\vec{p}} - \tilde{\vec{D}}\vec{x} \rVert_2^2 + \alpha \lVert \vec{x} \rVert_1$\\
        \textit{Dictionary Update:} \\
        $\hat{\vec{D}} = \arg\min_\vec{D} \lVert \tilde{\vec{p}} - \tilde{\vec{D}}\vec{x} \rVert_2^2 + \beta \sum_{l=1}^L \lVert \vec{H}(\frac{\omega_l}{c})\vec{d}_l \rVert_2^2$
        }
\end{algorithm}

Unlike the dictionary learning method~\cite{hahmannSpatialReconstructionSound2021}, our approach is \textit{zero-shot}, as it does not involve any pre-training on datasets. In fact, the algorithm is provided only with the observed pressure field $\tilde{\vec{p}}$; both the dictionary and the weights are jointly learned using this information alone, along with the Helmholtz regularization, that influences the dictionary update step (while the sparse coding remains unchanged).
Finally, once the dictionary $\vec{D}$ and weights $\vec{x}$ are learned, the sound field in the whole evaluation region is reconstructed via~\eqref{eq:reconstruction_equation}.
\section{Evaluation}\label{sec:evaluation}
We conduct an evaluation campaign based on experimental data to assess the performance of the proposed method. The experiments rely on the dataset introduced in~\cite{hahmann2021a}: the measurements are taken in a classroom with dimensions $[6.61, 9.45, 2.97]\si{\meter}$ having non-flat walls, an acoustically absorbing ceiling and containing scattering objects (e.g., desks and chairs), with $T_{60} = \SI{0.5}{\s}$. A single source is located in a room corner and impulse responses are recorded on a square grid of $69\times69$ points with $h=\SI{2.5}{\cm}$ spacing, over a planar aperture with size $\SI{1.7}{\meter}\times \SI{1.7}{\meter}$, centered at position $[3.78, 1.90, 1.86]\si{\meter}$ (further details can be found in~\cite{hahmannSpatialReconstructionSound2021}).

\begin{figure*}
     \centering
     \begin{subfigure}{\columnwidth}
         \centering
\begin{tikzpicture}

\definecolor{darkcyan0114178}{RGB}{0,114,178}
\definecolor{darkcyan0158115}{RGB}{0,158,115}
\definecolor{darkgray176}{RGB}{176,176,176}
\definecolor{lightsteelblue179205227}{RGB}{179,205,227}
\definecolor{navajowhite253220160}{RGB}{253,220,160}
\definecolor{orange2301590}{RGB}{230,159,0}
\definecolor{palegreen178226161}{RGB}{178,226,161}

\begin{axis}[%
    xlabel={Frequency},%
    x unit= \hertz,%
    ylabel={$\mathrm{NMSE}$},%
    y unit = \decibel, %
    axis x line=bottom,%
    axis y line=left, %
    grid, %
    height=3.6cm, %
    width=0.96\columnwidth, %
    enlarge y limits=0.1,%
    style={font=\normalsize},%
    legend columns=3,%
    legend style={at={(0.68,0.05)},anchor=south,font=\scriptsize},%
    log ticks with fixed point,
    legend cell align={left}
]
\path [fill=lightsteelblue179205227, fill opacity=0.5]
(axis cs:600,-1.02586543272262)
--(axis cs:600,-1.11210817862851)
--(axis cs:800,-0.419631637218398)
--(axis cs:1000,-0.163517804333596)
--(axis cs:1200,-0.103093600595758)
--(axis cs:1200,-0.0825795433564842)
--(axis cs:1200,-0.0825795433564842)
--(axis cs:1000,-0.138768909014421)
--(axis cs:800,-0.399665774370873)
--(axis cs:600,-1.02586543272262)
--cycle;

\path [fill=navajowhite253220160, fill opacity=0.5]
(axis cs:600,-2.95503877177157)
--(axis cs:600,-4.15792908178023)
--(axis cs:800,-2.07802312675997)
--(axis cs:1000,-1.12302178419249)
--(axis cs:1200,-0.424010704330933)
--(axis cs:1200,-0.299623253439104)
--(axis cs:1200,-0.299623253439104)
--(axis cs:1000,-0.820476235369814)
--(axis cs:800,-1.86812031479983)
--(axis cs:600,-2.95503877177157)
--cycle;

\path [fill=palegreen178226161, fill opacity=0.5]
(axis cs:600,-4.13485506928467)
--(axis cs:600,-4.22691960712335)
--(axis cs:800,-2.1274317421363)
--(axis cs:1000,-1.2641920306832)
--(axis cs:1200,-0.822359034042658)
--(axis cs:1200,-0.724372837221463)
--(axis cs:1200,-0.724372837221463)
--(axis cs:1000,-1.15209206340937)
--(axis cs:800,-2.03921611235887)
--(axis cs:600,-4.13485506928467)
--cycle;

\addplot [line width=0.35mm, orange2301590]
table {%
600 -3.5564839267759
800 -1.9730717207799
1000 -0.971749009781153
1200 -0.361816978885018
};
\addlegendentry{Proposed};
\addplot [line width=0.35mm, dashed, darkcyan0158115]
table {%
600 -4.18088733820401
800 -2.08332392724759
1000 -1.20814204704628
1200 -0.77336593563206
};
\addlegendentry{OLDL};
\addplot [line width=0.35mm, dotted, darkcyan0114178]
table {%
600 -1.06898680567557
800 -0.409648705794636
1000 -0.151143356674008
1200 -0.092836571976121
};
\addlegendentry{BL};
\end{axis}

\end{tikzpicture}
         \label{fig:nmse_freq}
     \end{subfigure}
     \hfill
     \begin{subfigure}{\columnwidth}
         \centering
\begin{tikzpicture}

\definecolor{darkcyan0114178}{RGB}{0,114,178}
\definecolor{darkcyan0158115}{RGB}{0,158,115}
\definecolor{darkgray176}{RGB}{176,176,176}
\definecolor{lightsteelblue179205227}{RGB}{179,205,227}
\definecolor{navajowhite253220160}{RGB}{253,220,160}
\definecolor{orange2301590}{RGB}{230,159,0}
\definecolor{palegreen178226161}{RGB}{178,226,161}

\begin{axis}[%
    xlabel={Frequency},%
    x unit= \hertz,%
    ylabel={$\mathrm{NCC}$},%
    axis x line=bottom,%
    axis y line=left, %
    grid, %
    height=3.6cm, %
    width=0.96\columnwidth, %
    enlarge y limits=0.1,%
    style={font=\normalsize},%
    legend columns=3,%
    legend style={at={(0.68,0.75)},anchor=south,font=\scriptsize},%
    log ticks with fixed point,
    legend cell align={left}
]
\path [fill=lightsteelblue179205227, fill opacity=0.5]
(axis cs:600,0.475261553310837)
--(axis cs:600,0.462178890629674)
--(axis cs:800,0.293898688414304)
--(axis cs:1000,0.184158503744898)
--(axis cs:1200,0.145016868826311)
--(axis cs:1200,0.16147231786918)
--(axis cs:1200,0.16147231786918)
--(axis cs:1000,0.200238952518634)
--(axis cs:800,0.30626984145468)
--(axis cs:600,0.475261553310837)
--cycle;

\path [fill=navajowhite253220160, fill opacity=0.5]
(axis cs:600,0.779714138660174)
--(axis cs:600,0.697094143556268)
--(axis cs:800,0.605778175633559)
--(axis cs:1000,0.469290647988052)
--(axis cs:1200,0.37683307180506)
--(axis cs:1200,0.401308850615651)
--(axis cs:1200,0.401308850615651)
--(axis cs:1000,0.50496349876174)
--(axis cs:800,0.625637195244898)
--(axis cs:600,0.779714138660174)
--cycle;

\path [fill=palegreen178226161, fill opacity=0.5]
(axis cs:600,0.806043380052111)
--(axis cs:600,0.795757137470892)
--(axis cs:800,0.636464916996293)
--(axis cs:1000,0.497996090787992)
--(axis cs:1200,0.398743505530483)
--(axis cs:1200,0.423748983728486)
--(axis cs:1200,0.423748983728486)
--(axis cs:1000,0.520333882234551)
--(axis cs:800,0.644884626210299)
--(axis cs:600,0.806043380052111)
--cycle;

\addplot [line width=0.35mm, orange2301590]
table {%
600 0.738404141108221
800 0.615707685439229
1000 0.487127073374896
1200 0.389070961210355
};
\addlegendentry{Proposed};

\addplot [line width=0.35mm, dashed, darkcyan0158115]
table {%
600 0.800900258761501
800 0.640674771603296
1000 0.509164986511272
1200 0.411246244629485
};
\addlegendentry{OLDL};

\addplot [line width=0.35mm, dotted, darkcyan0114178]
table {%
600 0.468720221970256
800 0.300084264934492
1000 0.192198728131766
1200 0.153244593347746
};
\addlegendentry{BL};
\end{axis}

\end{tikzpicture}
         \label{fig:ncc_freq}
     \end{subfigure}
     \caption{Average (solid line) and standard deviation (shaded area) of the $\operatorname{NMSE}$ and $\operatorname{NCC}$ over the 5 folds, with respect to frequency. The number of available measurements is $M = 50$.}
     \label{fig:nmse_ncc_freq}
\end{figure*}
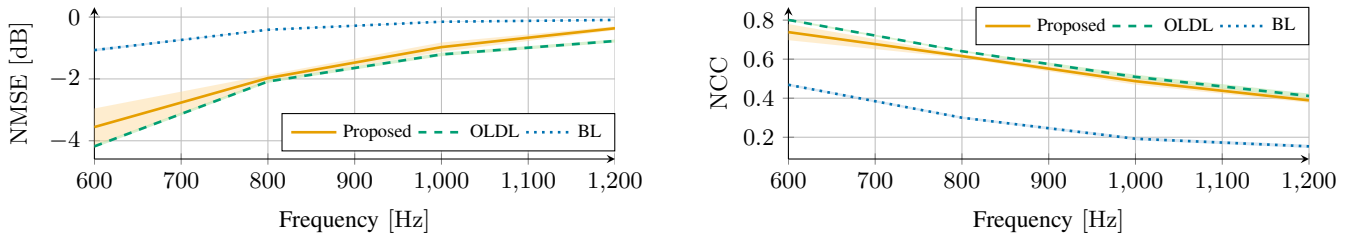

\begin{figure*}
     \centering
     \begin{subfigure}{\columnwidth}
         \centering
\begin{tikzpicture}

\definecolor{darkcyan0114178}{RGB}{0,114,178}
\definecolor{darkcyan0158115}{RGB}{0,158,115}
\definecolor{darkgray176}{RGB}{176,176,176}
\definecolor{lightsteelblue179205227}{RGB}{179,205,227}
\definecolor{navajowhite253220160}{RGB}{253,220,160}
\definecolor{orange2301590}{RGB}{230,159,0}
\definecolor{palegreen178226161}{RGB}{178,226,161}

\begin{axis}[%
    xlabel={$M$},%
    ylabel={$\mathrm{NMSE}$},%
    y unit = \decibel, %
    axis x line=bottom,%
    axis y line=left, %
    grid, %
    height=3.6cm, %
    width=0.96\columnwidth, %
    enlarge y limits=0.1,%
    style={font=\normalsize},%
    legend columns=3,%
    legend style={at={(0.33,0.02)},anchor=south,font=\scriptsize},%
    log ticks with fixed point,
    legend cell align={left}
]
\path [fill=lightsteelblue179205227, fill opacity=0.5]
(axis cs:10,-0.0517252415259143)
--(axis cs:10,-0.117575481435764)
--(axis cs:20,-0.394163017098493)
--(axis cs:30,-0.705654950229247)
--(axis cs:40,-0.885180296784235)
--(axis cs:50,-1.11210817862851)
--(axis cs:50,-1.02586543272262)
--(axis cs:50,-1.02586543272262)
--(axis cs:40,-0.796275676699301)
--(axis cs:30,-0.532724108147853)
--(axis cs:20,-0.286892615487153)
--(axis cs:10,-0.0517252415259143)
--cycle;

\path [fill=navajowhite253220160, fill opacity=0.5]
(axis cs:10,-0.151876053447396)
--(axis cs:10,-0.374269569934846)
--(axis cs:20,-1.47036873213514)
--(axis cs:30,-2.23229287195469)
--(axis cs:40,-3.06630490421484)
--(axis cs:50,-4.15792908178023)
--(axis cs:50,-2.95503877177157)
--(axis cs:50,-2.95503877177157)
--(axis cs:40,-2.53139469697151)
--(axis cs:30,-1.41626543689885)
--(axis cs:20,-0.596439395330238)
--(axis cs:10,-0.151876053447396)
--cycle;

\path [fill=palegreen178226161, fill opacity=0.5]
(axis cs:10,-0.563895269271879)
--(axis cs:10,-0.709882462102654)
--(axis cs:20,-1.56526715210578)
--(axis cs:30,-2.31804779643446)
--(axis cs:40,-3.1628787075834)
--(axis cs:50,-4.22691960712335)
--(axis cs:50,-4.13485506928467)
--(axis cs:50,-4.13485506928467)
--(axis cs:40,-3.09796936536303)
--(axis cs:30,-2.21718045730909)
--(axis cs:20,-1.39750615784864)
--(axis cs:10,-0.563895269271879)
--cycle;

\addplot [line width=0.35mm, orange2301590]
table {%
10 -0.263072811691121
20 -1.03340406373269
30 -1.82427915442677
40 -2.79884980059318
50 -3.5564839267759
};
\addlegendentry{Proposed};
\addplot [line width=0.35mm, dashed, darkcyan0158115]
table {%
10 -0.636888865687266
20 -1.48138665497721
30 -2.26761412687178
40 -3.13042403647321
50 -4.18088733820401
};
\addlegendentry{OLDL};
\addplot [line width=0.35mm, dotted, darkcyan0114178]
table {%
10 -0.0846503614808391
20 -0.340527816292823
30 -0.61918952918855
40 -0.840727986741768
50 -1.06898680567557
};
\addlegendentry{BL};
\end{axis}

\end{tikzpicture}
         \label{fig:nmse_mics}
     \end{subfigure}
     \hfill
     \begin{subfigure}{\columnwidth}
         \centering
\begin{tikzpicture}

\definecolor{darkcyan0114178}{RGB}{0,114,178}
\definecolor{darkcyan0158115}{RGB}{0,158,115}
\definecolor{darkgray176}{RGB}{176,176,176}
\definecolor{lightsteelblue179205227}{RGB}{179,205,227}
\definecolor{navajowhite253220160}{RGB}{253,220,160}
\definecolor{orange2301590}{RGB}{230,159,0}
\definecolor{palegreen178226161}{RGB}{178,226,161}

\begin{axis}[%
    xlabel={$M$},%
    ylabel={$\mathrm{NCC}$},%
    axis x line=bottom,%
    axis y line=left, %
    grid, %
    height=3.6cm, %
    width=0.96\columnwidth, %
    enlarge y limits=0.1,%
    style={font=\normalsize},%
    legend columns=3,%
    legend style={at={(0.72,0.02)},anchor=south,font=\scriptsize},%
    log ticks with fixed point,
    legend cell align={left}
]
\path [fill=lightsteelblue179205227, fill opacity=0.5]
(axis cs:10,0.20622241946931)
--(axis cs:10,0.17008453148263)
--(axis cs:20,0.26727042743573)
--(axis cs:30,0.345648442040581)
--(axis cs:40,0.415060513281182)
--(axis cs:50,0.462178890629674)
--(axis cs:50,0.475261553310837)
--(axis cs:50,0.475261553310837)
--(axis cs:40,0.432458286013497)
--(axis cs:30,0.397471845133327)
--(axis cs:20,0.309007483466751)
--(axis cs:10,0.20622241946931)
--cycle;

\path [fill=navajowhite253220160, fill opacity=0.5]
(axis cs:10,0.289082080665042)
--(axis cs:10,0.234850604376084)
--(axis cs:20,0.439317356551035)
--(axis cs:30,0.551034177979028)
--(axis cs:40,0.667243487460321)
--(axis cs:50,0.697094143556268)
--(axis cs:50,0.779714138660174)
--(axis cs:50,0.779714138660174)
--(axis cs:40,0.706372997264549)
--(axis cs:30,0.639761034825854)
--(axis cs:20,0.53634790927953)
--(axis cs:10,0.289082080665042)
--cycle;

\path [fill=palegreen178226161, fill opacity=0.5]
(axis cs:10,0.388935295729726)
--(axis cs:10,0.341836208102077)
--(axis cs:20,0.534441693981932)
--(axis cs:30,0.654118730850221)
--(axis cs:40,0.732302346446115)
--(axis cs:50,0.795757137470892)
--(axis cs:50,0.806043380052111)
--(axis cs:50,0.806043380052111)
--(axis cs:40,0.741397086522537)
--(axis cs:30,0.662162584999911)
--(axis cs:20,0.559213528224105)
--(axis cs:10,0.388935295729726)
--cycle;

\addplot [line width=0.35mm, orange2301590]
table {%
10 0.261966342520563
20 0.487832632915283
30 0.595397606402441
40 0.686808242362435
50 0.738404141108221
};
\addlegendentry{Proposed};
\addplot [line width=0.35mm, dashed, darkcyan0158115]
table {%
10 0.365385751915902
20 0.546827611103018
30 0.658140657925066
40 0.736849716484326
50 0.800900258761501
};
\addlegendentry{OLDL};
\addplot [line width=0.35mm, dotted, darkcyan0114178]
table {%
10 0.18815347547597
20 0.28813895545124
30 0.371560143586954
40 0.423759399647339
50 0.468720221970256
};
\addlegendentry{BL};
\end{axis}

\end{tikzpicture}
         \label{fig:ncc_mics}
     \end{subfigure}
     \caption{Average (solid line) and standard deviation (shaded area) of the $\operatorname{NMSE}$ and $\operatorname{NCC}$ over the 5 folds, with respect to the number of available measurements $M$, in the frequency range $[500,700]\si{\hertz}$.}
     \label{fig:nmse_ncc_mics}
\end{figure*}
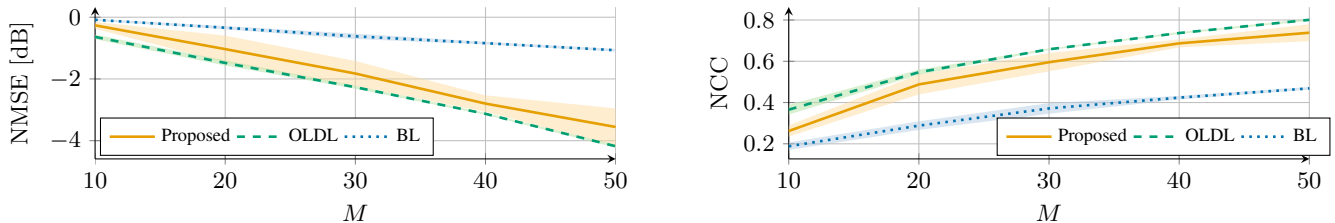

\begin{figure*}
    \centering
    \includegraphics[width=0.9\textwidth]{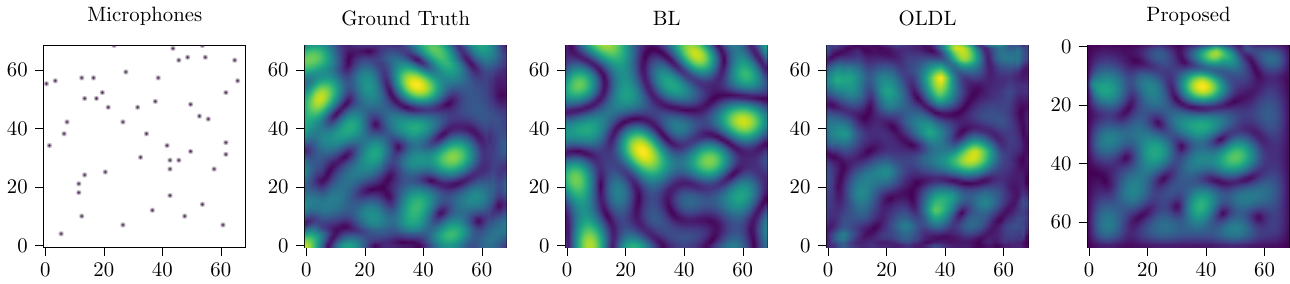}
     \caption{Magnitude of the sound field at $f=\SI{600}{\hertz}$ on the $69\times 69$ point grid, reconstructed using the BL, OLDL and proposed method and compared with the measured ground truth. The reconstruction is performed using $M=50$ microphones, depicted on the leftmost plot.}
     \label{fig:sound_field_plot}
\end{figure*}

To assess the performance of the proposed method, we compare it with a baseline (BL) and to the state-of-the-art dictionary learning sound field reconstruction method~\cite{hahmannSpatialReconstructionSound2021} (OLDL). The baseline adopts a fixed dictionary inspired by the \textit{synthesized} dictionary used in~\cite{hahmannSpatialReconstructionSound2021}. The frequency-dependent BL dictionary $\vec{D}_\text{bl}$ is built using $L=21$ atoms, drawn from (frequency-dependent) multivariate normal distributions with covariance matrix $\vec{\Sigma} \in \mathbb{R}^{N\times N}$, whose entry at position $s,t$ is obtained by~\cite{hahmannSpatialReconstructionSound2021} 
\begin{equation}
    \Sigma_{s,t} = j_0(k\lVert \vec{r}_s - \vec{r}_t \rVert_2)\,,
\end{equation}
where $j_0$ is the zero-order spherical Bessel function.
Similarly to the proposed method, each atom $\vec{d}_l$ is generated using spherical Bessel functions at frequency $\omega_l$ (thus, each atom represents a different frequency).

The OLDL method is trained using the procedure described in~\cite{hahmannSpatialReconstructionSound2021}, based on analyzing the sound field in local patches and pre-training the dictionary $\vec{D}_\text{oldl}$ in a training room\footnote{The code implementing this method is available at~\texttt{\url{https://github.com/manvhah/local_soundfield_reconstruction}}}. 

We implement the proposed method in Python\footnote{Source code will be made available upon acceptance.}, and use the CVXPY~\cite{diamond2016cvxpy} library to solve the optimization problem. For each experiment, we use a dictionary of $L=21$ atoms and a range of frequencies of $\SI{200}{\hertz}$ for the Helmholtz regularization: this results in atoms modeling frequencies spaced by $\SI{10}{\hertz}$. Preliminary analysis showed that increasing the number of atoms does not lead to an improved performance. We set the maximum number of iterations $I=40$ for solving the optimization problem, and set the hyperparameters $\alpha=1$ and $\beta=0.1$, obtained via a grid search. To emphasize the benefit of the physics-informed learning procedure, the dictionary $\vec{D}$ is initialized using the baseline dictionary $\vec{D}_\text{bl}$.

\subsection{Metrics}
We evaluate the performance of the proposed method based on two metrics: the normalized mean-squared error (NMSE) in \SI{}{\decibel} scale, defined as
\begin{equation}\label{eq:nmse}
    \operatorname{NMSE} = 10\log_{10} \left( \frac{\lVert \vec{p} - \hat{\vec{p}} \rVert_2^2}{\lVert \vec{p} \rVert_2^2} \right) \,,
\end{equation}
where $\hat{\vec{p}} \in \mathbb{C}^{N^2}$ denotes the estimated pressure at the $N^2$ points of the evaluation grid, and the normalized cross-correlation (NCC), defined as
\begin{equation}\label{eq:ncc}
    \operatorname{NCC} = \frac{\lvert \hat{\vec{p}}^\mathrm{H}\vec{p}\rvert}{\lVert \hat{\vec{p}}\rVert_2^2 \lVert \vec{p}\rVert_2^2} \,,
\end{equation}
where $(\cdot)^\mathrm{H}$ indicates the Hermitian transpose. Specifically, $\operatorname{NCC}$ produces a value ranging from \(-1\) to \(1\), where \(1\), \(0\), and \(-1\) correspond to perfect similarity, no correlation, and perfect anti-correlation, respectively \cite{morena2024reconstruction}.

\subsection{Results}
To ensure a robust evaluation, we examine the behavior of the proposed, BL and OLDL \cite{hahmannSpatialReconstructionSound2021} methods across varying frequencies and different numbers of available measurements $M$. All experiments are repeated 5 times, using randomly extracted configurations of measurement microphones $M$, and the average and standard deviation of the results over the 5 folds are reported for each experiment.

We first consider a fixed number of measurement microphones $M=50$ and analyze the performance of the proposed method across frequencies, considering the range from $\SI{600}{\hertz}$ to $\SI{1200}{\hertz}$, chosen to allow for a direct comparison with~\cite{hahmannSpatialReconstructionSound2021}. The entire frequency range is analyzed in the sub-bands $[500,700]\si{\hertz}, [700,900]\si{\hertz}, [900,1100]\si{\hertz}, [1100,1300]\si{\hertz}$. For each sub-band, a different dictionary of $L$ atoms is generated for the BL, OLDL and proposed methods, and the reconstruction is performed at $81$ frequency (i.e., every $\SI{2.5}{\hertz}$) within the same sub-band; we report the average of the results across each frequency band. Note that, since the atoms represent frequencies spaced by $\SI{10}{\hertz}$, the reconstruction is thus performed also at non-directly-modeled frequencies, where an interpolation between dictionary atoms is implicitly performed during the reconstruction.

In Fig.~\ref{fig:nmse_ncc_freq} we report the $\operatorname{NMSE}$ and the $\operatorname{NCC}$ across the considered frequency range. The performance of all methods degrades as frequency increases: for a fixed number of microphones $M$ this is to be expected, as the number of measurements per wavelength decreases as frequency increases, making the problem more challenging. The OLDL and the proposed method both outperform the BL in the entire frequency range, achieving a comparable performance. Nevertheless, while OLDL is a supervised learning method requiring pre-training on an extensive dataset, the proposed method works in a zero-shot setting, where the only data used for the reconstruction consists of the recording of the 50 measurement microphones in the evaluation room. We remark that, in the proposed method, $\vec{D}$ is intialized using the BL dictionary $\vec{D}_\text{bl}$: this ensures that the improvement of the proposed method over the BL directly results from the physically-informed learning strategy.

We then consider the frequency range $[500,700]\si{\hertz}$ and analyze the performance when a decreasing number of microphones is available. Fig.~\ref{fig:nmse_ncc_mics} shows that, as expected, the performance of all methods degrades as the number of microphones decreases. As in Fig.~\ref{fig:nmse_ncc_freq}, both OLDL and the proposed method outperform BL for all considered $M$, with OLDL achieving slightly better scores. 

We finally show in Fig.~\ref{fig:sound_field_plot} the reconstructed sound field at $f=\SI{600}{\hertz}$ using $M=50$ measurement microphones using the three methods, as compared to the ground truth measured field. This analysis confirms that the proposed method, similarly to OLDL, is able to achieve closer match to the measured field than BL.

The results underline the potential of the proposed dictionary learning approach for sound field reconstruction, and the effectiveness of imposing a physical constraint on the dictionary atoms as an alternative to pre-training the dictionary on an extensive training set. The proposed approach, in fact, guarantees results comparable with the state-of-the-art supervised dictionary learning method~\cite{hahmannSpatialReconstructionSound2021} without the need for any training data.

\section{Conclusion}\label{sec:conclusion}
In this paper, we proposed a physics-informed dictionary learning approach for sound field reconstruction. By introducing a regularization term in the learning objective that encourages the dictionary atoms to satisfy the Helmholtz equation, the proposed approach is able to learn a wideband representation of the sound field in a room in a zero-shot setting, requiring neither any prior knowledge on the acoustic scene nor the need for pre-training the dictionary using extensive training data. Preliminary studies indicate that the proposed approach achieves a reconstruction performance comparable with state-of-the-art supervised dictionary learning methods, without needing any training data and having the advantage of retrieving a physically meaningful dictionary. Future work will focus on comparing the proposed method with other non-dictionary-based techniques and on investigating the introduction of prior knowledge on the analyzed acoustic scene in the algorithm.

\section*{Acknowledgment}
We would like to express our gratitude to Manuel Hahmann for sharing code and data that helped in the development of this work.
\bibliographystyle{ieeetr}
\bibliography{refs}

\end{document}